\documentclass[10pt,reqno]{amsart}

\usepackage{amsmath,amsfonts,amsthm,amssymb,amsxtra}
\usepackage{amsxtra, amssymb, mathrsfs, pstricks}

\newtheorem{theorem}{Theorem}[section]
\newtheorem{proposition}[theorem]{Proposition}

\newtheorem{lemma}[theorem]{Lemma}
\newtheorem{corollary}[theorem]{Corollary}

\theoremstyle{definition}

\newtheorem{assumption}[theorem]{Assumption}

\theoremstyle{remark}
\newtheorem{remark}[theorem]{\bf Remark}

\numberwithin{equation}{section}

\newcommand{\A}{\mathcal{A}}

\newcommand{\h}{\mathcal{H}}
\newcommand{\Q}{\mathcal{Q}}

\newcommand{\pd}{\partial}

\newcommand{\N}{\mathbb{N}}

\newcommand{\R}{\mathbb{R}}

\newcommand{\Z}{\mathbb{Z}}

\begin{document}

\title[Large time behavior of magnetic heat kernels]
{Large time behavior of the heat kernel of two-dimensional magnetic
Schr\"odinger operators}

\author {Hynek Kova\v{r}\'{\i}k}

\address {Hynek Kova\v{r}\'{\i}k, Dipartimento di Matematica, Politecnico di Torino}

\email {Hynek.Kovarik@polito.it}


\date{\today}

\begin {abstract}
We study the heat semigroup generated by two-dimensional
Schr\"odinger operators with compactly supported magnetic field. We
show that if the field is radial, then the large time behavior of
the associated heat kernel is determined by its total flux.  We also
establish some on-diagonal heat kernel estimates and discuss their
applications for solutions to the heat equation. An exact formula
for the heat kernel, and for its large time asymptotic, is derived
in the case of the Aharonov-Bohm magnetic field.
\end{abstract}

\maketitle




\section{Introduction}

The Hamiltonian of a charged quantum particle in $\R^d$ interacting with
a magnetic field $B$ is given formally by the differential operator
\begin{equation} \label{ham-general}
H_B = ( i \nabla +A)^2
\end{equation}
in $L^2(\R^d)$, where $A$ is the vector potential of the magnetic
field; $B = \mbox{rot} A$ (for $d=2,3$). The object of our interest
in the present paper is the integral kernel $e^{-tH_B}(x,y)$ of the
heat semigroup generated by $H_B$. In particular, we are interested
in the dependence of $e^{-tH_B}(x,y)$ on $t$. A well known
semiclassical result, \cite{erd, mats} says that under certain
conditions on $B$ we have
\begin{equation} \label{short-time}
\lim_{t\to 0}\, t^{\frac d2}\, e^{-tH_B}(x,x) = (4 \pi)^{-d/2} .
\end{equation}
In other words, the leading term of $e^{-tH_B}(x,x)$ in the short
time limit is not affected by the magnetic field. However, the
situation changes in the large time limit, where the diagonal
element of the heat kernel decays exponentially fast provided the
size of the magnetic field is bounded from below by a positive
constant, \cite{erd2,mal}. More precisely, the estimate
\begin{equation} \label{large-time}
\lim_{t\to\infty} \frac 1t\, \log \| e^{-tH_{L B}}\|_{L^1\to L^\infty} \,
\leq \, -C_L \, L\, \min_{x\in \R^d}\, \|B(x)\|
\end{equation}
holds true with $C_L = 1 +o(1)$ as $L\to\infty$, see \cite{erd2}.
From the Mehler formula for the heat kernel of the two-dimensional
Schr\"odinger operator with a constant magnetic field, see
\cite[Sect.II.4-6]{sim}, it follows that the factor $\min_{x\in
  \R^d}\, \|B(x)\|$
in \eqref{large-time} cannot be improved.
Later, a uniform pointwise bound on the two-dimensional magnetic heat
kernel in the form
\begin{equation} \label{lt}
\| e^{-tH_{B}}\|_{L^1\to L^\infty}\, \leq \, \frac{B_0}{4\pi
\sinh(\frac{B_0 t}{2})}\, ,\quad B_0=  \min_{x\in \R^2}\, |B(x)|,
\quad t>0, \quad d=2
\end{equation}
was obtained in \cite{lt} under the assumption that $B_0>0$. This
bound is the best possible since there is equality for $B=B_0$. The
latter follows again by the Mehler formula.

In this paper we focus on the the case $d=2$ and address the
following question: what is the large time behavior of
$e^{-tH_B}(x,y)$ when $B(x)$ is of compact support? Note that for a
compactly supported magnetic field we have $B_0 = \min_{x\in \R^2}\,
|B(x)|=0$ in \eqref{large-time} and \eqref{lt}. This of course
reflects the fact that $\inf \text{spect} (H_B)=0$ and therefore no
exponential decay of the heat kernel is possible.

On the other hand, Laptev and Weidl showed in \cite{lw} that under
certain conditions on $B$ the operator $H_B$ satisfies a Hardy type
inequality
\begin{equation} \label{lw}
H_B \, \geq\, \frac{C_B}{1+|x|^2}
\end{equation}
in the sense of quadratic forms on $H^1(\R^2)$, see also
\cite{timo}. Inequality \eqref{lw} implies that $H_B$ is a
subcritical operator. The criticality theory then suggests that the
integral
\begin{equation} \label{green}
\int_0^\infty e^{-tH_B}(x,y)\, dt
\end{equation}
should be finite for all $x\neq y$. Hence in the limit of large
times the magnetic heat kernel $e^{-tH_B}(x,y)$ should behave
differently than the heat kernel of the usual Laplace operator in
$\R^2$. Our motivation is to find out how exactly the large time
behavior of $e^{-tH_B}(x,y)$ depends on the magnetic field.

One of our main results, Theorem \ref{main}, shows that for radially
symmetric and weak magnetic fields the time decay of
$e^{-tH_B}(x,y)$ is completely determined by the total flux of the
magnetic field. The key point of the proof is to show that
$e^{-tH_B}(x,y)$ is asymptotically (as $t\to\infty$) equivalent to
the heat kernel of certain two-dimensional Schr\"odinger operator
with positive potential, see \eqref{reference}. In section
\ref{sect-l2} we establish some pointwise and $L^p-$estimates on the
magnetic semigroup $e^{-tH_B}$ in terms of the distance between the
total flux and the set of integers, see Theorem \ref{cor} and
Proposition \ref{prop-l2}. One of the main technical tools used in
the proofs is Lemma \ref{bessel}, in which we derive a formula for
the heat semigroup of certain family of one-dimensional
Schr\"odinger operators. This also enables us to write an explicit
expression for the heat kernel of the Aharonov-Bohm operator,
Proposition \ref{expl}, and consequently to calculate its exact
large time asymptotic, see Theorem \ref{ab-thm}.

\begin{remark} \label{gauge}
Let us make a brief remark on the properties of
the heat kernel under gauge transformations. It is a matter of fact
that the vector potential $A$ is not uniquely determined by the
magnetic field $B$. However, if $\mbox{rot} A =\mbox{rot}\tilde A=
B\in C(\R^2,\R)$, then there exists a scalar field $\phi$ such that
$\tilde A= A +\nabla \phi$. So the respective Hamiltonians $H_B$ and
$\tilde{H}_B$ are unitarily equivalent; $ \tilde{H}_B = e^{i \phi}\,
H_B\, e^{-i \phi}$, and their heat kernels are linked through the
equation
$$
e^{- t \tilde{H}_B}(x,y) = e^{i (\phi(x)-\phi(y))}\, e^{-tH_B}(x,y).
$$
Hence changing the gauge does not change the time dependence of the
heat kernel. In other words, the decay rate in time is gauge
invariant.
\end{remark}

\section{Preliminaries}

\noindent Given two functions $f_1,\, f_2$ on a set $\Omega$ we will
use the notation $f_1 \simeq f_2$ to indicate that there exist
positive constants $c,C$ such that the inequalities $c\, f_1 \leq
f_2 \leq C\, f_1$ hold on $\Omega$. Accordingly, the notation
$f_1(t,x) \simeq f_2(t,x)$ as $t\to\infty$ means that $f_1 \simeq
f_2$ holds for all $t$ large enough. Moreover, given two points
$x,y\in\R^2$, we will often use the polar coordinate representation
$e^{-tH_B}(x,y) = e^{-tH_B}(r,r',\theta,\theta')$ of the heat kernel
which corresponds to the identification $x=r(\cos\theta,
\sin\theta)$ and $y= r'(\cos\theta', \sin\theta')$. Finally, we
denote $\R_+ =(0,\infty)$ and $\dot\R_+=[0,\infty)$. We will need
the following hypotheses.

\begin{assumption} \label{ass-field}
Let the magnetic field be given as $B(|x|),\, x\in \R^2$, where
$B:\dot\R_+\to \R$ is a continuous function with the support
contained in some interval $[0,R], \, R<\infty$.
\end{assumption}

\noindent We introduce a vector potential
$A:\dot\R_+\times[0,2\pi)\to\R^2$ which in polar coordinates
$(r,\theta)$ reads as follows
$$
A(r,\theta) = a(r)\, (-\sin\theta,\, \cos\theta), \quad a(r) = \frac
1r  \int_0^r\! B(t)\, t\, dt.
$$
Then $A$ generates the magnetic field $B$. Hamiltonian $H_B$ is
associated with the closed quadratic form
\begin{equation} \label{q-form}
Q_B[u] = \int_0^\infty\! \int_0^{2\pi} \left(|\pd_r u|^2+ r^{-2}
|i\, \pd_\theta u+b(r) u|^2\right) r\, dr d\theta, \quad u\in
H^1(\R_+\times(0,2\pi)),
\end{equation}
where
$$
b(r) = r a(r)=\int_0^r\! B(t)\, t\, dt = \frac{1}{2\pi}
\int_{\{|x|\leq r\}} B(|x|)\, dx
$$
is the flux of the magnetic field through the disc of radius $r$
centered in the origin. Moreover, we denote by $\alpha$ the total
flux of the magnetic field through the plane. By assumption
\ref{ass-field} we have
\begin{equation} \label{flux}
b(r) = \alpha\qquad \forall\, r>R.
\end{equation}
By expanding a given function $u\in L^2(\R_+\times (0,2\pi))$ into a
Fourier series with respect to the basis $\{e^{i
  m\theta}\}_{m\in\Z}$ of $L^2((0,2\pi))$, we obtain
a direct sum decomposition
\begin{equation}
L^2(\R^2) = \sum_{m\in\Z} \oplus \, \mathcal{L}_m,
\end{equation}
where $\mathcal{L}_m= \left\{g\in L^2(\R^2)\, : \,  g(x)= f(r)\,
e^{i m\theta} \, a.e., \, \int_0^\infty |f(r)|^2\,  r\,
dr<\infty\right\}$. Since the magnetic field $B$ is radial, the
operator $H_B$ can be decomposed accordingly to the direct sum
\begin{equation} \label{sum-gen}
H_B =   \sum_{m\in\Z}  \oplus \left( h_m \otimes\mbox{id}\right)
\Pi_m,
\end{equation}
where $h_m$ are operators generated by the closures, in $L^2(\R_+, r
dr)$, of the quadratic forms
\begin{equation} \label{qm}
Q_m [f] = \int_0^\infty\, \left(f'^2+\frac{(b(r)+m)^2}{r^2}\,
f^2\right)\, r\, dr
\end{equation}
defined initially on $C_0^\infty(0,\infty)$, and $\Pi_m:L^2(\R^2)\to
\mathcal{L}_m$ is the projector acting as
$$
(\Pi_m u)(r,\theta) = \frac{1}{2\pi}\, \int_0^{2\pi}\,
e^{im(\theta-\theta')}\, u(r,\theta')\, d\theta'.
$$
Note that $\Pi_m$ commutes with $h_m\otimes \mbox{id}$. Hence the
integral kernel of $e^{-tH_B}$ splits as follows:
\begin{equation} \label{hk-gen}
e^{-tH_B}(x,y)= \frac{1}{2\pi}\, \sum_{m\in\Z}\, p_m(r,r',t)\,
e^{im(\theta-\theta')}\, .
\end{equation}
Here $p_m(r,r',t)$ denotes $e^{-th_m}(r,r')$ which is real and
positive for all $m\in\Z$. The idea behind the proof of Theorem
\ref{main} is to show that if the magnetic flux is small enough,
then the large time behavior of $e^{-tH_B}(x,y)$ is determined by
the contribution from $m=0$ in \eqref{hk-gen}.


\section{Heat kernel of the reduced operators}
\label{aux-sec}

In this section we will study the heat kernels $p_m(r,r',t)$. First
we prove a result which allows us to quantify the large time
behavior of $p_0(r,r',t)$. To this end we consider an auxiliary
Schr\"odinger operator
\begin{equation} \label{reference}
\A =-\Delta + \frac{b^2(|x|)}{|x|^2} \quad \text{in\, \, }
L^2(\R^2).
\end{equation}
The operator $\A$ can be defined
in the usual way through the corresponding closed quadratic form
$$
Q_\A[f] = \int_{\R^2} \Big (|\nabla f|^2 + f^2\,  \frac{b^2(|x|)}{|x|^2} \Big)\,
dx, \qquad f\in D(Q_\A) = H^1(\R^2).
$$

\begin{lemma} \label{gr-state}
Assume that $b(\cdot)$ is not identically zero. Then there exists a
positive radial function $h\in C^2(\R^2)$ such that $\A\, h =0$.
Moreover, any such function satisfies \begin{equation} \label{gr-state-eq}
h(x) = h(|x|) \, \simeq\,  \left\{
\begin{array}{l@{\quad}cr}
|x|^{|\alpha|} & \text{if\, \,} & \alpha \neq 0 \, , \\
1+ |\log |x|| & \, \, \text{if \, }  & \alpha =0.
\end{array}
\right. \qquad  |x| >R.
\end{equation}
Finally, there exist positive constants $C$ and $c$ such that the
heat kernel of $\A$ admits for all $x,y\in \R^2$ and all $t>0$ the
following estimate,
\begin{equation} \label{heat-kernel-2d}
e^{-t \A}(x,y) \, \simeq \, C\, \frac{h(x)\, h(y)}{t\,
h(|x|+\sqrt{t})\, h(|y|+\sqrt{t})}\, \, e^{-c\, \frac{|x-y|^2}{t}}.
\end{equation}
\end{lemma}
\noindent Here we use, with a slight abuse of notation, the same symbol for
the function $h$ on $\R^2$ and for its natural identification on
$\dot\R_+$.

\begin{proof}
Since $b(|x|)= \alpha$ for $|x|>R$, the spectrum of $\A$ coincides
with the positive half-line $[0,\infty)$. Hence by the
Allegretto-Piepenbrink theorem, see e.g. \cite{mp}, there exists a
positive solution $u$ to the equation $\A\, u=0$. Since the
potential term $b^2(|x|)/|x|^2$ in $\A$ is H\"older continuous, see
assumption \ref{ass-field}, the elliptic regularity ensures that
$u\in C^2(\R^2)$. The radial function $h$ given by
$$
h(|x|) = \int_0^{2\pi} u(|x|,\theta)\, d\theta,
$$
then also satisfies $\A\, h=0$ and for $|x|> R$ we have
\begin{align}
h(x) = h(|x|)  & = a\, |x|^{|\alpha|} +b\, |x|^{-|\alpha|}, \quad \alpha\neq 0
\label{nonzero}\\
h(x) = h(|x|)  & = c +d\, |\log |x||,  \qquad \, \, \, \,  \,   \alpha = 0.
\label{zero}
\end{align}
The positivity of $h$ implies that $a\geq 0,\, d\geq 0$.  On the
other hand, $h$ satisfies $r (r\, h'(r))' = h(r)\,
b^2(r)$ with $r=|x|$  and therefore it is easy to see that $h$ is 
an increasing function of $r$. 
This means that $a>0,\, d>0$.  A straightforward verification
now shows that the manifold $\R^2$ equipped with the Lebesgue
measure and the function $h$ satisfy hypothesis of
\cite[Thm.10.10.(i)]{grig05}. The latter yields the heat kernel
estimate \eqref{heat-kernel-2d}.
\end{proof}

\begin{corollary} \label{groundstate}
There exists a positive radial function $h\in C^2(\R^2)$ such that
$H_B\, h =0$. Moreover, if  $b(\cdot)$ is not identically zero, then
any such function satisfies \eqref{gr-state-eq}.
\end{corollary}

\begin{proof}
This follows from Lemma \ref{gr-state} and the fact that the
operators $H_B$ and $\A$ coincide on the set of radial functions.
\end{proof}

\noindent In order to control the terms in \eqref{hk-gen} with
$m\neq 0$ we will make use of Lemma \ref{bessel} below which gives
an explicit formula for the heat semigroup generated by the
operators which are associated with the quadratic form
\begin{equation}
\Q_\beta [u] = \int_0^\infty \left(u'^2+\frac{\beta^2}{r^2}\,
u^2\right)\, r\, dr ,\qquad \beta \in\R
\end{equation}
defined on $C_0^\infty(\R_+)$. This form is closable, see e.g.
\cite[Sec.1.8]{da}, and its closure generates a self-adjoint
operator $\h_\beta$ in $L^2(\R_+, r dr)$. By the Beurling-Deny
criteria $\h_\beta$ generates on $L^2(\R_+, r dr)$ a symmetric
submarkovian semigroup $e^{-t \h_\beta}$. Let $e^{-t
\h_\beta}(r,r')$ be its integral kernel.

\begin{lemma} \label{bessel} Let $\h_\beta$ be the operator in
  $L^2(\R_+, r dr)$
associated with closure of the form $\Q_\beta$. Then for all $r,r'
\in \R_+$ and all $t>0$ it holds
\begin{equation} \label{bessel-eq}
e^{-t\, \h_\beta}(r,r') = \frac{1}{2 t}\, \,
I_{|\beta|}\left(\frac{r\, r'}{2t}\right)\,
e^{-\frac{r^2+r'^2}{4t}},
\end{equation}
where $I_{|\beta|}$ is the modified Bessel function of the first kind,
see e.g.
\cite[Chap.9]{as}.
\end{lemma}

\begin{proof}
Consider the operators
\begin{equation} \label{lbeta}
L_\beta = U\, \h_\beta\, U^{-1} \qquad \text{in \, \, \, } L^2(\R_+,
dr),
\end{equation}
where $U: L^2(\R_+, r\, dr)\to L^2(\R_+, dr)$ is a unitary mapping
acting as $(U f)(r) = r^{1/2} f(r)$. Note that $L_\beta$ is subject
to Dirichlet boundary condition at $0$ and that it coincides with
the Friedrichs extension of the differential operator
$$
-\frac{d^2}{dr^2}\, +\,  \frac{\beta^2-\frac 14}{r^2}
$$
defined on $C_0^\infty(\R_+)$. Denote by $D(L_\beta)$ the domain of
$L_\beta$. Now let $\lambda$ be a complex number from some fixed
neighborhood of $\R_+$. A straightforward calculation using the
standard technique of the Sturm-Liouville theory shows that the
integral kernel of the resolvent operator $(L_\beta-\lambda)^{-1}$
for $r<r'$ is given as follows
\begin{align*}
(L_\beta-\lambda)^{-1}(r,r') & = \frac{\pi i}{2}\,  \sqrt{r r' }\,
\,J_{|\beta|}(r\sqrt{\lambda}\, ) \left(
J_{|\beta|}(r'\sqrt{\lambda}\, ) + i\, Y_{|\beta|}(r
'\sqrt{\lambda}\, ) \right) , \qquad {\rm Im\, }\lambda
>0
\\
(L_\beta-\lambda)^{-1}(r,r') & = -\frac{\pi i}{2}\, \sqrt{r r'}\,
\,J_{|\beta|}(r\sqrt{\lambda}\, ) \left(\, J_{|\beta|}(r'
\sqrt{\lambda}\, )  - i\, Y_{|\beta|}(r '\sqrt{\lambda}\, ) \right)
 ,\quad {\rm Im\, }\lambda <0,
\end{align*}
where $J_{|\beta|}$ and $Y_{|\beta|}$ are the Bessel functions of
the first and second kind respectively. Next we introduce the
function $g(r,\lambda) =\sqrt{r}\, J_{|\beta|}(r\sqrt{\lambda})$,
and note that $L_\beta\, g= \lambda\, g$ and $g(0,\lambda)=0$. Hence
the Weyl-Titchmarsh-Kodaira Theorem, see \cite[Chap.13]{ds}, says
that
\begin{equation} \label{diag}
W_\beta\, L_\beta\, W_\beta^{-1}\, \varphi(\lambda) = \lambda\,
\varphi(\lambda), \quad \varphi\in W_\beta(D(L_\beta)),
\end{equation}
where the mapping $W_{\beta}$ and its inverse $W_\beta^{-1}$ given
by
\begin{equation} \label{fb}
(W_\beta\, u)(\lambda) = \int_0^\infty u(r)\sqrt{r}\,
J_{|\beta|}(r\sqrt{\lambda})\, dr, \quad (W_\beta^{-1} \varphi)(r) =
\int_0^\infty \varphi(\lambda)\sqrt{r}\,
J_{|\beta|}(r\sqrt{\lambda})\, \frac{d\lambda}{2}
\end{equation}
defined initially on $C_0^\infty(\R_+)$ extend to unitary operators
from $L^2(\R_+)$ onto itself.
Given $f\in C_0^\infty(\R_+)$, in view of \eqref{diag} we then get
\begin{align}
\left( e^{-t L_\beta}\, f\right)(r) & = \left( W_\beta^{-1}\, e^{-t
\lambda}\, W_\beta\, f\right)(r) = \int_0^\infty \sqrt{r r'}\,
\int_0^\infty e^{-t \lambda}  J_{|\beta|}(r\sqrt{\lambda})\,
J_{|\beta|}(r'\sqrt{\lambda})\,  \frac{d\lambda}{2}\, f(r')
\, dr' \nonumber \\
& = \frac{1}{2t}\, \int_0^\infty \sqrt{r r'}\, \,
I_{|\beta|}\left(\frac{r\, r'}{2t}\right)\,
e^{-\frac{r^2+r'^2}{4t}}\, f(r')\, dr', \label{hk-lbeta}
\end{align}
where we have used Fubini's theorem to switch the order of
integration and \cite[Eq.8.11(23)]{erde} to evaluate the
$\lambda-$integral. Moreover, since $\sqrt{r'}\, \, I_{|\beta|}(r
r'/2t)\, e^{-\frac{r^2+r'^2}{4t}}\in L^2(\R_+)$ for all $r,t>0$, see
\cite[Chap.9.7]{as}, identity \eqref{hk-lbeta} extends by density to
all $f\in L^2(\R_+)$. Hence
\begin{equation}
e^{-t L_\beta}(r,r') := \frac{1}{2t}\, \sqrt{r r'}\, \,
I_{|\beta|}\left(\frac{r\, r'}{2t}\right)\, e^{-\frac{r^2+r'^2}{4t}}
\end{equation}
is the integral kernel of $e^{-t L_\beta}$, and by \eqref{lbeta} we
conclude that
\begin{equation} \label{p0-eq}
e^{-t \h_\beta}(r,r')= \frac{1}{\sqrt{r\, r'}}\, \, e^{-t
L_\beta}(r,r') = \frac{1}{2 t}\, \, I_{|\beta|}\left(\frac{r\,
r'}{2t}\right)\, e^{-\frac{r^2+r'^2}{4t}}.
\end{equation}
\end{proof}

\begin{lemma} \label{compare}
Let $|\alpha| <1$. Then for all $x,y\in \R^2$ it holds
\begin{equation} \label{compare-eq}
\lim_{t\to\infty}\, e^{-t \A}(x,y)\, (p_0(|x|,|y|,t))^{-1}  = 1.
\end{equation}
\end{lemma}

\begin{proof}
Operator $\A$ admits the decomposition
\begin{equation}
\A =  \sum_{m\in\Z}  \oplus \left( \A_m \otimes\mbox{id}\right)
\Pi_m,
\end{equation}
where $\A_m$ are operators in $L^2(\R_+, r\, dr)$ generated by the
closures of the quadratic forms
$$
a_m [f] = \int_0^\infty\, \left(f'^2+\frac{b(r)^2+m^2}{r^2}\,
f^2\right)\, r\, dr
$$
defined on $C_0^\infty(0,\infty)$. Note that $\A_0=h_0$ and hence
\begin{equation} \label{sum-A}
e^{-t \A}(x,y) = p_0(|x|,|y|,t) + \sum_{m\neq 0}\, e^{-t
\A_m}(|x|,|y|)\, e^{im( \theta-\theta')}.
\end{equation}
In order to estimate the sum on the right hand side of the last
equation, we note that by the Trotter
product formula
\begin{equation} \label{upperb}
e^{-t \A_m}(r,r')\, \leq \, e^{-t \h_m}(r,r') \qquad \forall \,
r,r'\in \R_+, \quad \forall\, m\neq 0,
\end{equation}
where $\h_m$ is the operator defined in Lemma \ref{bessel}. By the
same Lemma we get
\begin{equation} \label{tz}
\big | \sum_{m\neq 0}\, e^{-t \A_m}(|x|,|y|)\, e^{im(
\theta-\theta')} \big | \,  \leq \,  C\, z\, \sum_{m\neq 0}\,
I_{|m|}(z)\, , \quad z := \frac{|x y|}{2t},
\end{equation}
where the constant $C$ depends on $x$ and $y$. Assume first that
$\alpha\neq 0$.
From the integral representation
\begin{equation} \label{int-repr}
I_\nu(z)= \frac{z^\nu}{2^\nu\, \Gamma(\nu+\frac 12)\Gamma(\frac 12)}\,
\int_{-1}^1\, (1-s^2)^{\nu-\frac 12}\, e^{zs}\, ds
\end{equation}
for $I_\nu$, see e.g. \cite[Chap.9]{as}, it is then easy to see that
\begin{align*}
& \limsup_{t\to\infty} \, t^{1+|\alpha|}\, \big | \sum_{m\neq 0}\,
e^{-t \A_m}(|x|,|y|)\, e^{im( \theta-\theta')}\, \big | \leq\, c\,
\limsup_{z\to 0} \, \sum_{n\geq 1}\, \frac{z^{n-|\alpha|}}{2^n\,
\Gamma(n+\frac 12)}
 \\
& \qquad \qquad  \leq \, c\, \limsup_{z\to 0} \, z^{1-|\alpha|}\,
\sum_{n\geq 1}\, \frac{1}{2^n\, \Gamma(n+\frac 12)} = 0,
\end{align*}
where $c'$ depends on $x$ and $y$. Since
$$
e^{-t \A}(x,y)  \, \simeq\, t^{-1-|\alpha|}\qquad t\to\infty, \quad
\alpha\neq 0,
$$
by Lemma \ref{gr-state}, we conclude from \eqref{sum-A} that
equation \eqref{compare-eq} holds true in the case $\alpha\neq 0$. On
the other hand,
if $\alpha =0$, then Lemma \ref{gr-state} gives
$$
e^{-t \A}(x,y)  \, \simeq\, t^{-1}\, (\log t)^{-2}\qquad t\to\infty,
\quad \alpha= 0.
$$
From  \eqref{tz} and \eqref{int-repr} we find
\begin{align*}
& \limsup_{t\to\infty} \, t\, (\log t)^{2}\, \big | \sum_{m\neq 0}\,
e^{-t \A_m}(|x|,|y|)\, e^{im( \theta-\theta')}\, \big |  \leq\, c\,
\limsup_{z\to 0}  \sum_{n\geq 1}\, \frac{(\log z)^2 \,
z^{n}}{2^{n}\, \Gamma(n+\frac 12)} =0.
\end{align*}
This proves \eqref{compare-eq} for $\alpha=0$.
\end{proof}


\section{Large time asymptotic of $e^{-tH_B}(x,y)$}
\label{main-sect}

Below we formulate our main result regarding the large time behavior
of the magnetic heat kernel $e^{-tH_B}(x,y)$. It shows that if the
magnetic field is sufficiently small, then the decay rate of
$e^{-tH_B}(x,y)$ is completely determined by the total flux
$\alpha$.

\begin{theorem} \label{main}
Let $B(x)$ satisfy assumption \ref{ass-field}  and
suppose that $|b(r)| < 1/2$ for all $r\in\R_+$.  Let $h\in C^2(\R^2)$ be a positive radial
function such that $H_B\, h=0$.
Then there exist constants $C$ and $c$ such that the inequalities
\begin{equation} \label{lim-nonzero}
c\, \leq \, \liminf_{t\to \infty}\, t^{1+|\alpha|}\, \,
\frac{e^{-tH_B}(x,y)}{h(x) h(y)} \,  \leq \, \limsup_{t\to \infty}\,
t^{1+|\alpha|}\, \, \frac{e^{-tH_B}(x,y)}{h(x) h(y) } \, \leq \, C,
\, \, \, \, \qquad \alpha\neq 0
\end{equation}
and
\begin{equation} \label{lim-zero}
c\, \leq \liminf_{t\to \infty}\, t\, (\log t)^2\, \, \frac{
e^{-tH_B}(x,y)}{h(x) h(y) }\,  \leq\,  \limsup_{t\to \infty}\, t \,
(\log t)^2\, \, \frac{e^{-tH_B}(x,y)}{h(x) h(y) } \, \leq \, C, \quad
\alpha= 0
\end{equation}
hold true for all $x,y \in \R^2$.
\end{theorem}

\begin{remark}
Similar connection between the large time
asymptotic of the heat kernel $e^{-t P}(x,y)$ and the ground state of
the corresponding generator is known when $P$ has an eigenvalue
at the bottom of its spectrum, see e.g. \cite{ck,pin,sim3}.
\end{remark}

\begin{remark}
Equation \eqref{lim-zero} shows that $e^{-tH_B}(x,y)$ is integrable
with respect to $t$ at infinity even if the total flux is zero. The
latter reflects the fact that $H_B$ satisfies a Hardy type
inequality also in this case, see \cite{timo}.
\end{remark}

\begin{proof}[Proof of Theorem \ref{main}]
The existence of the ground state $h$ is guaranteed by Corollary \ref{groundstate}.
By Lemma \ref{gr-state} it suffices to show that
\begin{equation} \label{equal-lim}
\lim_{t\to\infty}\, \frac{e^{-t \A}(x,y)}{e^{-t H_B}(x,y)} = 1 \qquad
\forall\, x , y \in\R^2.
\end{equation}
Let $\alpha\neq 0$. By assumption we have $(b(r)+m)^2 \geq m^2/4$
for all $m\neq 0$ and all $r\in\R_+$. Hence the Trotter product
formula gives
$$
p_m(r,r',t) \, \leq\, e^{-t\, \h_{m/2}}(r,r')\qquad \forall\, \, r,
r'\in \R_+, \, \, t>0, \,   m\neq0.
$$
With the notation of equation \eqref{tz} we get from
\eqref{int-repr} and Lemma \ref{bessel}
\begin{align*}
\limsup_{t\to\infty} \, t^{1+|\alpha|}\, \sum_{m\neq 0}\,
p_m(r,r',t) \leq\, c \, \limsup_{z\to 0}\, z\, \sum_{m\neq 0}\,
I_{|m/2|}(z) \, \leq  c\, \lim_{z\to 0} \, \sum_{n\geq 1}\,
\frac{z^{\frac n2-|\alpha|}}{2^{n/2}\, \Gamma((n+1)/2)} =0,
\end{align*}
where we have used the fact that $|\alpha| < 1/2$. In view of
equations \eqref{hk-gen}, \eqref{heat-kernel-2d} and Lemma
\ref{compare}, this proves  \eqref{equal-lim}. If $\alpha =0$, we
obtain in the same way as above
\begin{align*}
\limsup_{t\to\infty} \, t\, (\log t)^2 \, \sum_{m\neq 0}\,
p_m(r,r',t) =0.
\end{align*}
Equation \eqref{equal-lim} thus holds also in this case.
\end{proof}

\noindent In the case $|\alpha| \geq 1/2$ we give an asymptotic
upper bound on the heat kernel.

\begin{proposition} \label{main-2}
Let $B(x)$ satisfy assumption \ref{ass-field}. Let $\varrho =
\min_{k\in\Z} |k+\alpha|$ be the distance between the flux $\alpha$ and the
set of integers. Then there exists a constant $C$ such that
\begin{equation} \label{lim-0.5}
\limsup_{t\to \infty}\, t^{1 +\varrho}\, | e^{-tH_B}(x,y)| \, \leq
\, C\, (1+|x|)^{\varrho}\, (1+|y|)^{\varrho }
\end{equation}
holds for all $x,y\in\R^2$.
\end{proposition}

\begin{proof}
We introduce the operators $T_m$ generated by the quadratic forms
$$
t_m [f] = \int_0^\infty\, \left(f'^2+ \Theta(r-R)\, \,
\frac{(b(r)+m)^2}{r^2}\, \, f^2\right)\, r\, dr,
$$
defined initially on $C_0(\R_+)$ and then closed in $L^2(\R_, r\,
dr)$. Here $\Theta(\cdot)$ denotes the Heaviside function. By the
Trotter product formula we have
\begin{equation} \label{upperb-m}
e^{-t h_m}(r,r')\, \leq \, e^{-t\, T_m}(r,r') \qquad \forall \,
r,r'\in \R_+, \quad  m\in\Z.
\end{equation}
In view of \eqref{flux} it follows that the functions $\psi_m\in
C^1(\R_+,\R_+)$, defined by
\begin{equation} \label{gm}
\psi_m(r) = 1, \quad  r < R, \quad \psi_m(r) =  \frac 12 \,
\left(\frac rR\right)^{\sigma_m} + \frac 12 \, \left(\frac R
r\right)^{\sigma_m}\, \quad r\geq R ,\quad \sigma_m=|\alpha+m|
\end{equation}
solve the Cauchy problems
$$
\Theta(r-R)\, \, \frac{(b(r)+m)^2}{r}\, \, \psi_m = (r\, \psi_m')',
\quad \psi_m(R)=1, \, \, \psi_m'(R)=0.
$$
The operators
$$
S_m = \psi_m^{-1}\, T_m\, \psi_m \qquad \text{in\, \,} L^2(\R_+,\,
\psi_m^2(r)\, r dr),
$$
are thus unitarily equivalent to $T_m$ and their heat kernels
satisfy
\begin{equation} \label{transf-heat}
e^{-t\, T_m}(r,r') = \psi_m(r)\, \psi_m(r')\, e^{-t S_m}(r,r').
\end{equation}
A direct calculation shows that $S_m$ is associated with the
quadratic form
$$
s_m [u]= t_m [u\, \psi_m]= \int_0^\infty (u')^2\, \psi_m^2\, r\, dr,
\qquad u\in D(s_m)= H^1(\R_+, \psi_m^2\, r\, dr).
$$
We now apply Theorem \ref{mazya} with $ \mu(x) = \nu(x)= x\,
\psi^2_m(x)$, $p=2$ and $q= (2+2\sigma_m)/\sigma_m$. Hence for each
$m$ there exists a constant $c_m$, such that
\begin{equation} \label{sobol}
s_m [u] \, \geq \,  c_m\, \Big(\int_0^\infty
u^{\frac{2+2\sigma_m}{\sigma_m}}\, \psi_m^2\, r\, dr
\Big)^{\frac{\sigma_m}{1+\sigma_m}} \qquad \forall\, u \in D(s_m).
\end{equation}
By the Beurling-Deny criteria, $S_m$ generates on $L^2(\R_+,
\psi_m(r) dr)$ a symmetric submarkovian semigroup $e^{-t S_m}$. This
allows us to apply \cite[Thm.2.4.2]{da}, see also \cite{var}, to
obtain
$$
\|e^{-t S_m}\|_{\infty,2}\, \leq \, C_m\, t^{-\frac{1+\sigma_m}{2}}
$$
for some constant $C_m$. By duality this implies that
$$
\sup_{r,r'}\, e^{-t S_m}(r,r') = \|e^{-t S_m}\|_{\infty,1}\, \leq\,
\|e^{-t S_m}\|^2_{\infty,2}\, \leq \, C_m^2\, t^{-1-\sigma_m} \qquad
\forall\, t>0.
$$
In view of equations \eqref{upperb-m} , \eqref{gm} and
\eqref{transf-heat} this yields
\begin{equation} \label{upperb-2}
e^{-t\, T_m}(r,r') \leq \, C_m^2\, \, t^{-1-\sigma_m}\,
(1+r)^{\sigma_m}\, (1+r')^{\sigma_m} \qquad t>0, \, \, r,r'\in\R_+.
\end{equation}
Now define $n_0:=\inf\{n\in\N\, :\, n > 2 \sup_{r>0}\, |b(r)| \}$.
From \eqref{upperb-m} and \eqref{upperb-2} we obtain
\begin{equation}
\limsup_{t\to\infty}\, t^{1+\varrho}\! \sum_{m=-n_0}^{n_0}
p_m(r,r',t) \leq \, C\, (1+r)^\varrho (1+r')^\varrho.
\end{equation}
To estimate the rest of the sum in \eqref{hk-gen} we note that
$$
(b(r)+m)^2 \geq \frac{m^2}{4} \qquad \forall\, \, r >0, \quad m\,
:\, |m| > n_0.
$$
Hence mimicking the arguments used in the proof of Theorem
\ref{main} it is easy to see that
$$
\limsup_{t\to\infty} \, t^{1+\varrho}\! \sum_{|m| >n_0} p_m(r,r',t)
=0.
$$
By \eqref{hk-gen} this completes the proof.
\end{proof}

\section{Heat kernel estimates}
\label{sect-l2}

In this section we use Theorem \ref{main} and Proposition
\ref{main-2} in order to prove certain point-wise heat kernel
estimates. We use the notation introduced in Proposition
\ref{main-2}, i.e. $\varrho = \min_{k\in\Z} |k+\alpha|$.

\begin{theorem} \label{cor}
Under assumption \ref{ass-field}
there exists a constant $C$ such that the inequality
\begin{equation} \label{hk-ub}
e^{-tH_B}(x,x) \,  \leq \, C\, \min\left\{ t^{-1},\,
(1+|x|)^{2\varrho }\, \, t^{-1-\varrho}\right\}
\end{equation}
holds for all $x\in\R^2$ and all $t>0$.
\end{theorem}

\begin{proof}
Adopting the notation of the proof of Theorem \ref{main-2}, it follows from \eqref{upperb-2}
that
$$
p_m(r,r,t) \leq C\, t^{-1-\sigma_m}\, (1+r)^{2 \, \sigma_m}  \qquad \forall\, m\in \{-n_0,\dots, n_0\}.
$$
On the other hand, the diamagnetic inequality
\begin{equation} \label{diamag}
\big | e^{-t H_B}(x,y)\big | \, \leq e^{t \Delta}(x,y) =
\frac{1}{4\pi t}\, e^{-\frac{|x-y|^2}{4t}} , \qquad \quad
x,y\in\R^2,\quad t>0,
\end{equation}
see e.g. \cite{ahs,hs,sim2}, clearly implies that $p_m(r,r,t)\leq
1/(2t)$ for all $m$. Hence
\begin{equation} \label{finite-m}
p_m(r,r,t) \leq C\, (t^{-1-\sigma_m}\, (1+r)^{2 \, \sigma_m} )^{\frac{\varrho}{\sigma_m}}\,
t^{-(1-\frac{\varrho}{\sigma_m})} = C\,  t^{-1-\varrho}\, (1+r)^{2 \, \varrho}
\quad |m|\leq n_0.
\end{equation}
Next we introduce the variable $z= \frac{|x|^2}{t}$. From the proof
of Theorem \ref{main} and from Lemma \ref{bessel} we get
\begin{align}
z^{-\varrho}\, \sum_{|m|>n_0} t\, p_m(|x|,|x|,t) &
\leq \, z^{-\varrho}\, \sum_{|m|>n_0} t\, e^{-t\,
  \h_{m/2}}(|x|,|x|) = c\, z^{-\varrho}\,e^{-z}\, \sum_{|m|>n_0}
I_{|m/2|}(z).  \label{pm-z}
\end{align}
On the other hand, inequality \eqref{diamag} shows that
$$
\sum_{|m|>n_0} t\, p_m(|x|,|x|,t) \leq  \sum_{m\in \Z} t\,
p_m(|x|,|x|,t) =2\pi\,  t\,  e^{-t H_B}(x,x) \leq \frac 12.
$$
This in combination with \eqref{pm-z} gives
\begin{align*}
\sup_{t,r>0}\, \frac{|x|^{2 \varrho}}{t^\varrho} \,
\sum_{|m|>n_0}  p_m(|x|,|x|,t)  \, & \leq \, c\, \max\Big\{
\sup_{z\leq 1} \, z^{-\varrho}\, \sum_{|m|>n_0}
I_{|m/2|}(z)\, , \, \sup_{z >1}\,  z^{-\varrho} \Big\} \leq C.
\end{align*}
Indeed, in view of \eqref{int-repr} the series $\sum_{|m|>n_0}
I_{|m/2|}(z)$ converges uniformly with respect to $z$ on $[0,1]$.
Hence $z^{-|\alpha|}\, \sum_{|m|>n_0} I_{|m/2|}(z)$ is continuous on
$(0,1]$ and since it tends to zero as $z\to 0$, see the proof of
Theorem \ref{main}, it is bounded. From equation \eqref{finite-m} we
thus get
\begin{align*}
\sum_{m\in\Z}  p_m(|x|,|x|,t)  \, & \leq \,   C\,
(1+|x|)^{2 \varrho}\, \, t^{-1-\varrho}, \qquad
\forall\, x\in\R^2, \quad t>0.
\end{align*}
The statement now follows by \eqref{hk-gen} and \eqref{diamag}.
\end{proof}


\noindent As a consequence of inequality \eqref{hk-ub} we get an
estimate on the norm of $e^{-t H_B}$ acting on certain weighted
$L^p$ spaces. To formulate our result we introduce the following
family of subspaces:
$$
L^p_\beta := \left\{ f\, :  \|f\|_{L^p_\beta} < \infty \right\},
\quad \|f\|_{L^p_\beta} := \left(\int_{\R^2} |f|^p\,
(1+|x|)^{\beta}\, d x \right)^{\frac 1p},\qquad \beta\in\R.
$$
We then have

\begin{proposition} \label{prop-l2}
Let assumptions \ref{ass-field} be satisfied. Assume that $p\in
[1,2]$ and let $q\in[2,\infty]$ be such that $\frac 1p +\frac 1q=1$.
Then for any $\beta
> 2+2\, \varrho$ there exists a constant $C=C(\varrho,\beta)$
such that
\begin{equation} \label{l2-estim} \| e^{-t
H_B}\|_{L^p_\beta\to L^q} \, \leq \, C\, \, t^{\frac{1-q}{q}
-\frac{\varrho}{2}} \qquad \forall\, t\geq 1.
\end{equation}
\end{proposition}

\begin{proof}
We use the shorthand $k(t,x,y)=e^{-t H_B}(x,y)$. Since $e^{-t H_B}$
is self-adjoint, we have $k(t,y,x) = \overline{k(t,x,y)}$. The
semigroup property of $e^{-t H_B}$ and the Cauchy-Schwarz inequality
then yield
\begin{align}
|k(2t,x,y)| & = \Big| \int_{\R^2} k(t,x,z)\, k(t,z,y)\, dz \Big|
\leq \Big( \int_{\R^2} |k(t,x,z)|^2\, dz\Big)^{\frac 12} \Big(
\int_{\R^2} |k(t,z,y)|^2\, dz\Big)^{\frac 12} \nonumber \\
& = \sqrt{k(2t,x,x)}\, \, \sqrt{k(2t,y,y)} \label{semigroup}.
\end{align}
This in combination with estimate \eqref{hk-ub} and diamagnetic
inequality \eqref{diamag} gives
\begin{equation} \label{interpolate}
|k(t,x,y)| \, \leq \, C\, t^{-1- \varrho\frac{\mu_1+\mu_2}{2}}\,
(1+|x|)^{\mu_1\, \varrho} (1+|y|)^{\mu_2\, \varrho},\quad \forall \,
\, \mu_1,\, \mu_2 \in [0,1].
\end{equation}
Now fix $f\in L^2(\R^2)$ and let $t\geq 1$. Chose $\mu_1=\mu_2=1$ in
\eqref{interpolate}. In view of \eqref{semigroup}, Cauchy-Schwarz
inequality and Fubuni's theorem we have
\begin{align}
\| e^{-t H_B}\, f \|^2_{L^2} & = \int_{\R^2} \Big| \int_{\R^2}
k(t,x, y) f(y)\, dy\big |^2 \, dx \, \leq \, \|f\|^2_{L^2_\beta}\,
\int_{\R^2} \int_{\R^2} |k(t,x, y)|^2 (1+|y|)^{-\beta} \, dy \, dx
\nonumber \\
& = \, \|f\|^2_{L^2_\beta}\,  \int_{\R^2} k(2t, y, y) \,
(1+|y|)^{-\beta}\, dy \, \leq \, C'\, \, t^{-1-\varrho}\, \,
\|f\|^2_{L^2_\beta}. \label{beta-norm}
\end{align}
This shows that
\begin{equation} \label{duality}
\| e^{-t H_B}\|_{L^2_{\beta}\to L^2} \, \leq \, C' \,
t^{-(1+\varrho)/2}.
\end{equation}
On the other hand, choosing $\mu_1= 0$ and $\mu_2=1$ in
\eqref{interpolate} it is easily seen that
\begin{equation} \label{l1-infty}
\| e^{-t H_B}\|_{L^1_{\beta}\to L^\infty} \, \leq \, C \,
t^{-1-\varrho/2}.
\end{equation}
Inequality \eqref{l2-estim} now follows from \eqref{duality},
\eqref{l1-infty} and the Riesz-Thorin interpolation theorem.
\end{proof}

\begin{remark}
In the absence of magnetic field we have
\begin{equation} \label{laplace}
\| e^{-t H_0}\|_{L^p_\beta \to L^q} = \| e^{t \Delta}\|_{L^p_\beta
\to L^q} \, \simeq \, C\, \, t^{\frac{1-q}{q}}\qquad \forall\, \beta
>2.
\end{equation}
Indeed, the upper bound in \eqref{laplace} follows by mimicking the
proof of Proposition \ref{prop-l2} with $k(t,x,y)$ replaced by $e^{t
\Delta}(x,y) = e^{-\frac{|x-y|^2}{4t}}/(4\pi t)$. This leads to
equations \eqref{duality} and \eqref{l1-infty} with $\varrho=0$. In
order to prove the lower bound in \eqref{laplace} let us consider
the solution of the heat equation with the initial data $f(x) =
e^{-|x|^2}$. An easy calculation gives
$$
u(t,x)= \left( e^{t \Delta} f\right)(x) = \frac{1}{1+4t}\, \,
e^{-\frac{|x|^2}{1+4t}} , \qquad \|u(t,\cdot)\|_{L^q} = c\,
(1+4t)^{\frac{1-q}{q}}.
$$
Proposition \ref{prop-l2} thus says that the $L^q$ norm of the
solution to the heat equation
$$
\partial_t u + H_B\, u =0, \qquad u(0,x)=f(x),
$$
decays faster (with respect to the case $B=0$), if we restrict the
initial data $f$ to a smaller subspace of $L^p(\R^2)$. Note also
that similar estimates were recently obtained, in the case
$p=q=2$, for the heat semigroup of Dirichlet-Laplace operator in
twisted waveguides; see \cite{kz}.
\end{remark}

\section{Example: The Aharonov-Bohm operator}

\noindent A natural question which arises from theorem \ref{main} is
whether the limit
\begin{equation} \label{limit-infty}
\lim_{t\to\infty} t^{1+|\alpha|}\, e^{-t H_B}(x,y)
\end{equation}
always exists and how it depends on $x$ and $y$. In this section we
calculate the limit \eqref{limit-infty} in the case of the so-called
Aharonov-Bohm magnetic field. This field is characterized by the
property that the flux $b(r)$ through a disc of radius $r$ is
constant. It is generated by the vector potential $A$ whose radial
and azimuthal components (in the polar coordinates) are given by
\begin{equation}
A(r,\theta) = (a_1(r,\theta),\, a_2(r)), \qquad a_1=0,\quad
a_2(r)= \left(0,\frac{\alpha}{r}\right) .
\end{equation}
The associated operator $(i\nabla + A)^2$ defined on
$C_0^\infty(\R^2\setminus\{0\})$ has deficiency indices $(2,2)$, see
\cite{at,pr}. We will consider the Hamiltonian $H_\alpha$ as its
Friedrichs extension. In other words, we define $H_\alpha$ as a non
negative self-adjoint operator in $L^2(\R^2)$ generated by the
closure of the quadratic form
$$
Q_\alpha[u] = \int_0^{2\pi}\! \int_0^\infty \big(|\pd_r u|^2
+r^{-2}\, |(-i\pd_\theta+\alpha)\, u|^2\big )\, r\, dr d\theta,
\quad u\in C_0^\infty((0,\infty)\times[0,2\pi)).
$$

\begin{proposition} \label{expl}
Let $t>0$ and $r,r'\in \R_+$. Then the heat kernel of the
Aharonov-Bohm Hamiltonian $H_\alpha$ is given by the absolutely
convergent series
\begin{equation} \label{general-eq}
e^{-t H_\alpha}(x,y)= \frac{1}{4\pi t}\, \,
e^{-\frac{r^2+r'^2}{4t}}\, \, \sum_{m\in\Z}\,
I_{|m+\alpha|}\left(\frac{r\, r' }{2t}\right)\,
e^{im(\theta-\theta')}.
\end{equation}
\end{proposition}

\begin{proof}
We note that
\begin{equation} \label{sum}
H_\alpha =  \sum_{m\in\Z}  \oplus \left( \h_{m+\alpha}\,
\otimes\mbox{id}\right) \Pi_m,
\end{equation}
where $\h_{m+\alpha}$ are the operators in $L^2(\R_+, r dr)$ defined
in Lemma \ref{bessel}. Hence
\begin{equation} \label{general}
e^{-t H_\alpha}(x,y) = \frac{1}{2\pi}\, \sum_{m\in\Z}\, e^{-t\,
\h_{m+\alpha}}(r,r')\, e^{im(\theta-\theta')}\, ,
\end{equation}
Equation \eqref{general-eq} now follows from Lemma \ref{bessel}. The
absolute convergence of the series is easily seen from the integral
representation of the Bessel function $I_\nu$, see equation
\eqref{int-repr}.
\end{proof}

\begin{remark}
For $\alpha\in\Z$ we get by \cite[Eq.9.6.33]{as}
\begin{equation} \label{alpha-integ}
e^{-t H_\alpha}(x,y) = \frac{1}{4\pi t}\, \,
e^{-\frac{r^2+r'^2}{4t}} \, e^{\frac{r r'}{2t}\,
\cos(\theta-\theta')} \, e^{i \alpha
 (\theta'-\theta)}
 = \frac{1}{4\pi t}\, \, e^{-\frac{|x-y|^2}{4t}} \, e^{i \alpha
 (\theta'-\theta)}.
\end{equation}
This reflects the well known fact that for integer values of the
flux the Aharonov-Bohm operator is unitarily equivalent to the
Laplacian in $L^2(\R^2)$ under the unitary mapping $f\mapsto
e^{-i\alpha\, \theta} f$ , see also Remark \ref{gauge}. Equation
\eqref{general-eq} also implies that it is no loss of generality if
we suppose that $\alpha \in [-1/2,\, 1/2]$.
\end{remark}

\begin{theorem} \label{ab-thm}
We have
\begin{align} \label{limit-global-1}
\lim_{t\to \infty}\, t^{1+ |\alpha|}\, e^{-t H_\alpha}(x,y) & =
\frac{1}{4\pi \Gamma(1+| \alpha|)}\, \left(\frac{ r\,
r'}{4}\right)^{|\alpha|} & \quad \text{if\, \, } |\alpha| < 1/2, \\
 \label{limit-global-2}
\lim_{t\to \infty}\, t^{\frac 32}\, e^{-t H_\alpha}(x,y) & =
\frac{1}{4\pi \Gamma(3/2)}\, \left(\frac{r\, r'}{4}\right)^{\frac
12}\, (1+e^{\mp i(\theta-\theta')}) & \quad \text{if\, \, } \alpha =
\pm 1/2.
\end{align}
\end{theorem}

\begin{proof}
From equation \eqref{bessel-eq} and the asymptotic behavior of
$I_\nu(z)$ for small $z$, see \cite[Chap.9]{as}, we get
\begin{align} \label{limit-0}
\lim_{t\to \infty}\, t^{1+{|m+\alpha|}}\,  e^{-t
\h_{m+\alpha}}(r,r') & = \frac{r^{|m+\alpha|}\,
r'^{|m+\alpha|}}{2^{2|m+\alpha|+1}\, \Gamma(1+|m+\alpha|)}\, .
\end{align}
Assume first that $|\alpha|< 1/2$. In view of \eqref{int-repr} we
obtain
$$
e^{-\frac{r^2+r'^2}{4t}}\,  I_{|m+\alpha|}\left(\frac{r\, r'
}{2t}\right) \leq t^{-|m+\alpha|}\, \frac{(r
r')^{|m+\alpha|}}{2^{|m+\alpha|}\, \Gamma(|m+\alpha|+\frac
12)\Gamma(\frac 12)}\, e^{\frac{-(r-r')^2}{4t}} .
$$
Since $ \inf_{m\neq 0} |m+\alpha| > |\alpha|$,  it follows that
$$
\lim_{t\to\infty} t^{|\alpha|}
e^{-\frac{r^2+r'^2}{4t}}\, \sum_{m\neq 0}\,
I_{|m+\alpha|}\left(\frac{r\, r' }{2t}\right)\, e^{im(\theta-\theta')} = 0,
$$
which, in combination with \eqref{limit-0}, proves equation
\eqref{limit-global-1}. The proof in the case $|\alpha| = 1/2$
follows the same line.
\end{proof}


\appendix

\section{}
\label{aux}

\noindent  For the reader's convenience, and also because equation
(13) of \cite[Sec.1.3.1]{m} contains a missprint, we recall below a
simplified version of \cite[Thm.1.3.1.3]{m}.

\begin{theorem}[\bf Maz'ya]  \label{mazya}
Let $1\leq p\leq q \leq\infty$ and let $\mu, \nu \in L^1(\R_+)$ be
nonnegative. Then the inequality
\begin{equation} \label{sob-maz}
\left( \int_0^\infty |u|^q\, \mu(x)\, dx\right)^{\frac 1q}\, \leq \,
C\, \left(\int_0^\infty |u'|^p\, \nu(x)\, dx \right)^{\frac 1 p}
\end{equation}
holds for all $u\in W^{1,p}(\R_+,\, \nu(x)dx)$ and some constant
$C$, independent of $u$, if and only if
\begin{equation} \label{condition}
\sup_{r>0}\,  \left(\int_0^r \mu(x)\, dx\right)^{\frac 1q}\,
\left(\int_r^\infty \nu(x)^{-\frac{1}{p-1}}\, dx
\right)^{\frac{p-1}{p}} < \infty.
\end{equation}
\end{theorem}


\section*{Acknowledgements}
I would like to thank Georgi Raikov for many helpful discussions.
The financial support of Centre Bernoulli, Ecole Polytechnique
F\'{e}d\'{e}rale de Lausanne, where a part of this work was done, is
gratefully acknowledged.


\bibliographystyle{amsalpha}

\end{document}